# PINNs for the Solution of the Hyperbolic Buckley-Leverett Problem with a Non-convex Flux Function

Waleed Diab and Mohammed Al Kobaisi

Department of Petroleum Engineering, Khalifa University of Science and Technology
P.O.Box 127788, Abu Dhabi, United Arab Emirates

___________________________________________________________________________

**Abstract**

The displacement of two immiscible fluids is a common problem in fluid flow in porous media. Such a problem can be posed as a partial differential equation (PDE) in what is commonly referred to as a Buckley-Leverett (B-L) problem. The B-L problem is a non-linear hyperbolic conservation law that is known to be notoriously difficult to solve using traditional numerical methods. Here, we address the forward hyperbolic B-L problem with a nonconvex flux function using physics-informed neural networks (PINNs). The contributions of this paper are twofold. First, we present a PINN approach to solve the hyperbolic B-L problem by embedding the Oleinik entropy condition into the neural network residual. We do not use a diffusion term (artificial viscosity) in the residual-loss, but we rely on the strong form of the PDE. Second, we use the Adam optimizer with residual-based adaptive refinement (RAR) algorithm to achieve an ultra-low loss without weighting. Our solution method can accurately capture the shock-front and produce an accurate overall solution. We report a $\mathbb{L}_2$ validation error of $\sim 2 \times 10^{-2}$ and a $\mathbb{L}_2$ loss of $\sim 1 \times 10^{-6}$. The proposed method does not require any additional regularization or weighting of losses to obtain such accurate solution.

___________________________________________________________________________

## 1. Introduction

Physics Informed Neural-Networks is an emerging approach in which it has become possible to embed domain knowledge into machine learning models. Specifically, it entertains the embedding of physical laws in deep neural networks (NNs) expressed in terms of partial differential equations (PDEs). Informing NNs with physical laws eliminates the need for large datasets and provides a theoretical prior for the machine learning model. PINNs can be viewed as an unsupervised learning method that can approximate solutions to PDEs with data only on the initial and boundary conditions, ushering a new class of PDE solvers that do not require numerical discretization. As such, PINNs is beginning to gain wide credence amongst the computational science research community.

In principle, PINNs assume that a deep neural network is a parametric function that itself can be the solution to a PDE if properly constrained. The key is to constrain the NN with a PDE residual, generally referred to as a soft penalty constraint. Thus, the optimized model favor solutions that adhere to the physics of the problem which reduces the reliance on data considerably. Contrary to data-driven approaches, PINNs can provide more robust models that can provide long-term, accurate, consistent, and more generalizable spatiotemporal predictions that are rooted in physics while leaving room for data assimilation [1]. Furthermore, the main advantage of PINNs over traditional discretization schemes is that it is mesh-free and therefore overcomes the curse of dimensionality [2]. In addition, it relies on the strong form of the PDE and Automatic Differentiation (AD) to compute derivatives exactly, thereby avoiding truncation errors.

___________________________

*Email addresses*: 100049411@ku.ac.ae (Waleed Diab), mohammed.alkobaisi@ku.ac.ae (Mohammed Al Kobaisi)

Another advantage of PINNs is that it can solve inverse problems as easily as it can solve forward problems [3]–[8].

PINNs in its current form was first introduced in [3], where the term was first coined, while earlier attempts to embed physical laws into NNs were made as early as the 1990s [9]–[12]. Recently, PINNs has been applied to various problems in physics and computational sciences, including but not limited to problems in fluid mechanics [13]–[16], quantum mechanics [3], [17], and fluid flow in porous media [18]–[23]. These works show the broad applicability of PINNs while achieving excellent predictive accuracy.

Applications in fluid flow in porous media are of particular interest to us. In this paper we solve the hyperbolic B-L problem with a nonconvex flux function; a scalar conservation law with a piecewise constant initial condition known as a Riemann type problem with a solution that involves a shock and a rarefaction wave. Recent studies addressing this problem using PINNs include [19] and [20]. A solution to the B-L problem by including an artificial viscosity (second-order derivative term) was first introduced in [19]. There are two main issues with this approach. First, and more importantly, the inclusion of a second-order term fundamentally changes the PDE from a hyperbolic PDE to a parabolic PDE which results in slight smearing of the shock-front and a reduction in the solution accuracy. Second, the PINN approach becomes significantly slower due to the computation of a second-order derivative of the NN. Another solution to the B-L problem was introduced in [20]. Their approach entails embedding the entropy and velocity constraints into the NN residual (i.e. constraining the loss landscape with more physics). This method solves the two issues associated with the vanishing viscosity approach. Although the solution method proposed in [20] captures the shock front quite nicely, it is necessary to have access to the NVIDIA SimNet package [24] and an NVIDIA GPU in order to replicate the results. In this work, we draw inspiration from the study in [20] and present a simpler and more accurate solution that does not require access to the SimNet package nor any additional regularization or weighting of losses to obtain such accurate solution.

The paper is organized as follows. In section 2, we define the nonlinear scalar conservation laws with emphasis on the B-L problem. Section 3 presents the details of PINNs and its mathematical formulation. Section 4 presents three solution methods to the B-L problem using PINNs with the nonconvex flux function including our proposed solution method. Finally, conclusions of our work are given in section 5.

## 2. Problem Definition: Nonlinear Scalar Conservation Laws

Scalar conservation laws take the following form:

$$u_t + f(u)_x = 0. \tag{1}$$

Here, $u(t, x)$ is the scalar conserved quantity that we solve for, $u_t$ is the partial derivative of the dependent quantity $u$ with respect to time $(t)$, $f(u)$ is a nonlinear flux function, and $f(u)_x$ is the partial derivative of the flux function with respect to position $(x)$. The solution to this problem involves a shock wave across which the solution is discontinuous. In our case, $f(u)$ is nonconvex ($f''(u)$ changes sign) resulting in a more complicated solution which involves both a shock and a rarefaction wave (compound wave).

A prominent example of these types of problems is the Buckley-Leverett (B-L) problem. The problem typically arises in the secondary recovery of oil from underground rock formations (porous medium). Water is injected into the porous medium through injection wells to displace additional amounts of unrecovered oil which is to be produced from other wells (producers). The classic two-phase displacement B-L problem is posed in a one-dimensional, homogeneous, and isotropic porous medium. The wetting phase, usually water ($w$), displaces the non-wetting phase, usually oil ($o$), and the two fluids are incompressible and immiscible.

Let $S_w(x,t)$ be the water saturation (corresponding to $u$ in Equation 1), and the oil saturation be $S_o(x,t) = 1 - S_w(x,t)$ as $0 \leq S_w \leq 1$. B-L assumes a 1-D horizontal porous medium, $x \in [0,1]$, and neglects capillarity. The problem also neglects volumetric flow terms, replacing them with a water inflow at $x = 0$. Formally:

$$S_w(x,0) = \begin{cases} 1 & \text{if } x < 0, \\ 0 & \text{if } x > 0. \end{cases} \tag{2}$$

The problem is governed by a coupled system of conservation equations supplemented by Darcy's equation, for a detailed derivation see e.g., [25] and [26]. The conservation equation for the water phase with associated initial and boundary conditions can be written as follows:

$$\frac{\partial S_w}{\partial t} + \frac{\partial f_w(S_w)}{\partial x} = 0, \quad S_w(x,0) = 0, \quad S_w(0,t) = 1. \tag{3}$$

Equation 3 is a Riemann type problem, a hyperbolic PDE with a piecewise constant initial condition (Equation 2) and a single jump discontinuity at $x = 0$. This discontinuity is a result of the sudden change in saturation between the initial water inflow and the current water saturation inside the domain. These types of problems are known to be notoriously difficult to solve using numerical methods. For a detailed and excellent discussion on Riemann type problems along with the known analytical and numerical solutions, the reader is referred to [27].

The flux function $f_w(S_w)$ is also called the water fractional flow and is defined as follows for the classical displacement case (irreducible water saturation $S_{w_{irr}} = 0$):

$$f_w(S_w) = \frac{1}{1 + \frac{\mu_o}{k_{rw}} \frac{k_{ro}}{\mu_w}} = \frac{S_w^2}{S_w^2 + \frac{(1-S_w)^2}{M}}, \tag{4}$$

where, $k_{rw}$ and $k_{ro}$ are the water and oil relative permeabilities given by the Brooks-Corey model [28], respectively. The model assumes a power-law relationship between saturation and relative permeability, with the exponents ($n_w = n_o = 2$), for water and oil, respectively. $M$ is the mobility ratio defined as the ratio of oil viscosity ($\mu_o$) to the water viscosity ($\mu_w$):

$$M = \frac{\mu_o}{\mu_w}. \tag{5}$$

In this work, $M$ is taken to be equal to 2 (oil is twice as viscous as water). The nonconvex fractional flow function is shown in Figure 1. This type (nonconvex) of fractional flow functions is the most commonly encountered in real-world applications. It is characterized by its S-shape with one inflection point, and the solution using it would have one discontinuity. Solution methods of the hyperbolic PDE (Equation 3) using PINNs will be discussed in the following section.

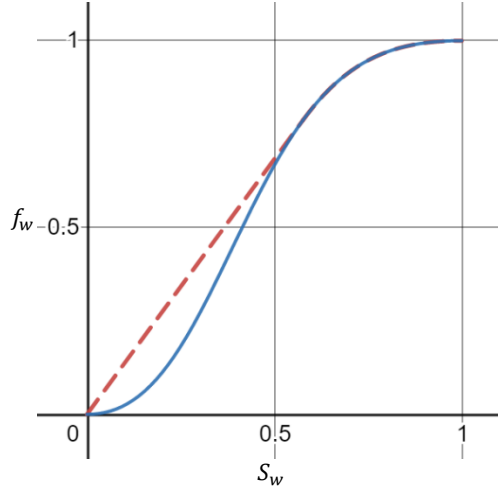

**Figure 1**: Nonconvex fractional flow function (blue) with an Oleinik entropy condition concave envelope.

## 3. Physics Informed Neural Networks (PINNs)

Let $u(x, t)$ be a function that solves a PDE. Building upon the assumption that deep NNs are universal function approximators, one can make use of this fact to find an approximate solution $\hat{u}(\mathbf{x}; \theta)$ using a deep NN. The simplest form of deep NNs, the feed-forward neural network (FNN), is a series of linear and nonlinear function transformations applied sequentially to an input ($\mathbf{x}$) to produce an output ($\hat{u}$). The nonlinearity is produced by applying a continuous and differentiable function to the output of each layer, generally referred to as the activation function ($\sigma$). In the same manner that $u_{x,t}$ can be found by differentiating $u$ with respect to its variables to build the PDE, $\hat{u}$ can be differentiated with respect to its inputs to build a surrogate of the PDE. This can be achieved through automatic differentiation (AD). AD computes derivatives by applying the chain rule repeatedly. AD capabilities can be leveraged through well-documented machine learning packages such as Tensorflow [29] and PyTorch [30].

In the interest of consistency in the literature, we follow [31] in the NN definition; we find their definitions appealing and mathematically robust. Let $\mathcal{N}^L(\mathbf{x}) : \mathbb{R}^{d_{in}} \to \mathbb{R}^{d_{out}}$ be an $(L-1)$-hidden layers NN, and $\mathbf{x} = (x_1, \dots, x_d; t_1, \dots, t_d)$ containing space and time coordinates be the input vector, with $N_\ell$ neurons in the $\ell$th layer ($N_0 = d_{in}$, $N_L = d_{out}$). We denote the weight matrix in the $\ell^{\text{th}}$ layer by $\mathbf{W}^\ell \in \mathbb{R}^{N_\ell \times N_{\ell-1}}$, and the bias vector $\mathbf{b}^\ell \in \mathbb{R}^{N_\ell}$. We refer to the combined set of all weight matrices and bias vectors as $\theta = \{\mathbf{W}^\ell, \mathbf{b}^\ell\}_{1 \le \ell \le L}$. We apply $\sigma$ elementwise where $\sigma$ is the hyperbolic tangent (tanh). We define the FNN as follows:

$$\text{input layer: } \mathcal{N}^0(\mathbf{x}) = \mathbf{x} \in \mathbb{R}^{d_{in}},$$

$$\text{hidden layer: } \mathcal{N}^\ell(\mathbf{x}) = \sigma\big(\mathbf{W}^\ell \mathcal{N}^{\ell-1}(\mathbf{x}) + \mathbf{b}^\ell\big) \in \mathbb{R}^{N_\ell} \text{ for } 1 \le \ell \le L-1 \quad (6)$$

$$\text{output layer: } \mathcal{N}^L(\mathbf{x}) = \mathbf{W}^L \mathcal{N}^{L-1}(\mathbf{x}) + \mathbf{b}^L \in \mathbb{R}^{d_{out}}$$

Although other deep NN architectures and activation functions have been proposed in the literature, we find the performance with a simple FNN and a hyperbolic tangent activation function satisfactory. Following the work in [3] for similar problems, and that of [19] and [20] on B-L problems, we use 8 hidden layers ($L = 9$) FNN with 20 neurons per layer ($N_\ell$=20).

To this end, the NN we built (Equation 6) is not physics informed. To inform the NN with physics as specified by the PDE we define the residual of the PDE ($r(x, t; \theta)$) to be the left-hand-side of Equation 1 and we replace $u$ by $\hat{u}$. In other words, we reconstruct the PDE using the NN and its derivatives as follows:

$$r := \hat{u}_t + f(\hat{u})_x. \tag{7}$$

The goal is to find a NN that satisfies both its input data (initial and boundary conditions) and the residual defined by Equation 7. Regardless of $\theta$, the NN must satisfy the PDE by construction, as the PDE is encoded in the NN in the form of prior knowledge. This approach is called physics-informed neural networks as was first proposed in [3]. The NNs defined by Equations 6 and 7 share the same trainable parameters $\theta$. To find the optimum $\theta^*$ is to train the NN. This is achieved by constructing an appropriate loss function and minimizing it using an appropriate optimizer. By leveraging AD, the loss function can be differentiated with respect to $\theta$, and minimized via a gradient-based optimizer. The loss function is defined as follows:

$$\mathcal{L}(\theta; N) = \lambda_{\hat{u}} \mathcal{L}_{\hat{u}}(\theta; N_{\hat{u}}) + \lambda_r \mathcal{L}_r(\theta; N_r), \tag{8}$$

where

$$\mathcal{L}_{\hat{u}}(\theta; N_{\hat{u}}) = \frac{1}{N_{\hat{u}}} \sum_{\mathbf{x} \in N_{\hat{u}}} \left\| \hat{u}(\mathbf{x}_{\hat{u}}^i) - u^i \right\|_2^2,$$

$$\mathcal{L}_r(\theta; N_r) = \frac{1}{N_r} \sum_{\mathbf{x} \in N_r} \left\| r(\mathbf{x}_r^i) \right\|_2^2.$$

In Equation 8, $\mathcal{L}_{\hat{u}}$ is the loss function on the initial and boundary data and $\{\mathbf{x}_{\hat{u}}^i, u^i\}_{i=1}^{N_{\hat{u}}}$ is the set of initial and boundary training data on $u(x, t)$; $\mathcal{L}_r$ imposes the physics as defined by the PDE and $\{\mathbf{x}_r^i\}_{i=1}^{N_r}$ is the set of interior (collocation) points where $r(x, t)$ is to be optimized. Based on the work presented in [19] and [20], 300 initial and boundary points ($N_{\hat{u}} = 300$) and 10,000 internal ($N_r = 10,000$) points were used to solve the B-L problem using PINNs. In this work, we use the same $N_{\hat{u}}$ and $N_r$ and the points are randomly sampled from the domain. $\lambda_{\hat{u}}$ and $\lambda_r$ are weighting parameters to control the contribution of each component to the loss. Here we set $\lambda_{\hat{u}} = \lambda_r = 1$.

Training the PINN model requires minimizing the loss function (Equation 8) with respect to $\theta$. The Adam algorithm [32] which relies on first-order derivatives and the limited-memory Broyden-Fletcher-Goldfarb-Shanno (L-BFGS) [33] which relies on second-order derivatives are among the optimizers best suited for the job since the loss function is highly nonlinear and nonconvex. Our findings suggest that the Adam algorithm achieves a lower loss when combined with the RAR algorithm (see Section 4) on the B-L problem with no diffusion. The complete PINN architecture is shown in Figure 2.

The authors in [31] proposed the residual-based adaptive refinement (RAR) algorithm to improve the distribution of residual points in the interior of the domain during training. After training for a pre-specified number of iterations (1 training loop), the algorithm finds the point where the mean residual is highest ($N_f^*$) and then adds $m$ points in the same location and goes through a training loop. The mean residual ($\varepsilon_r$) is computed as follows:

$$\varepsilon_r = \frac{1}{(N_f + mN_f^*)} \sum_{N_f} \|f(\mathbf{x}; \theta)\|. \qquad (9)$$

This process stops when the mean residual is below a predetermined threshold ($\varepsilon_0$). This would also serve as a weighting of the loss function in favor of the residual at locations where it is still high. The number of residual points to be added ($m$) per training loop can be considered as a hyperparameter and we find the best performance on the B-L problem with $m = 4$ points.

We explore three solution methods in Section 4. The hyperparameters for all the solution methods are listed in Table 1. All NNs are of the same design (8 layers deep, 20 neurons wide, $N_\ell = 20$) with a hyperbolic tangent activation function in all deep layers. We use a total of $N_{\hat{u}} = 300$ randomly distributed initial and boundary data and $N_f = 10{,}000$ interior points sampled from the domain $x \in [0,1], t \in [0,1]$ according to a Latin Hypercube Sampling strategy to enforce the PDE (Figure 3(a)). This sampling strategy was shown to produce better convergence with PINNs [3]. In addition, the weights and biases are initialized using the Xavier initialization scheme [34]. The initialized solution surface is shown in Figure 3(b).

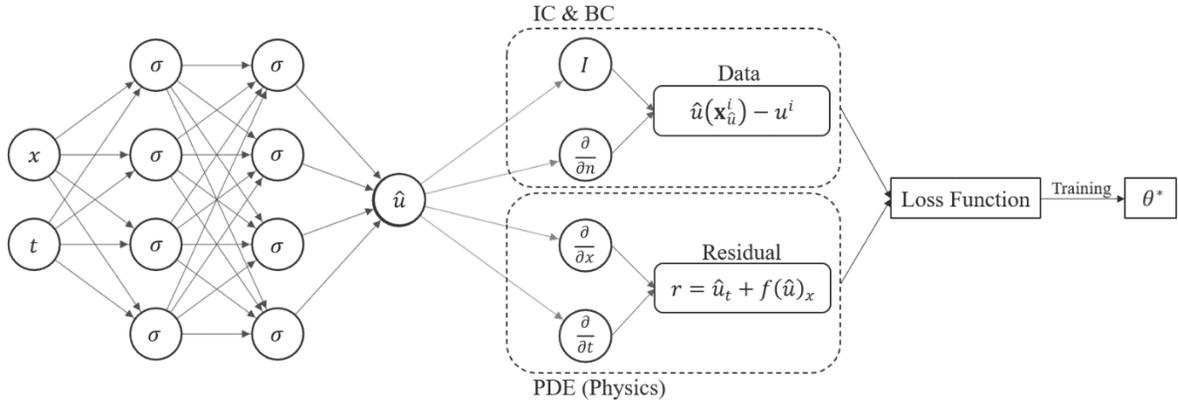

**Figure 2**: PINN architecture for the solution of the hyperbolic B-L problem.

**Table 1:** Hyperparameters used in the solution of the B-L problem in sections 4.1-4.3. Note that the optimizer L-BFGS does not require a learning rate.

| Solution Method | L | $N_\ell$ | $N_{\hat{u}}$ | $N_f$ | Optimizer | $\epsilon$ | Learning rate | Epochs* | # of Added Residual Points per Training loop |
|---|---|---|---|---|---|---|---|---|---|
| 4.1 | 9 | 20 | 300 | 10,000 | L-BFGS | 0 | - | - | 0 |
| 4.2 | | | | | L-BFGS | 0.0025 | - | - | 0 |
| 4.3 | | | | | L-BFGS | - | - | 2,800 | 4 |
| | | | | | Adam | - | 0.01 | 15,000 | 4 |

*The number of epochs prior to RAR algorithm.

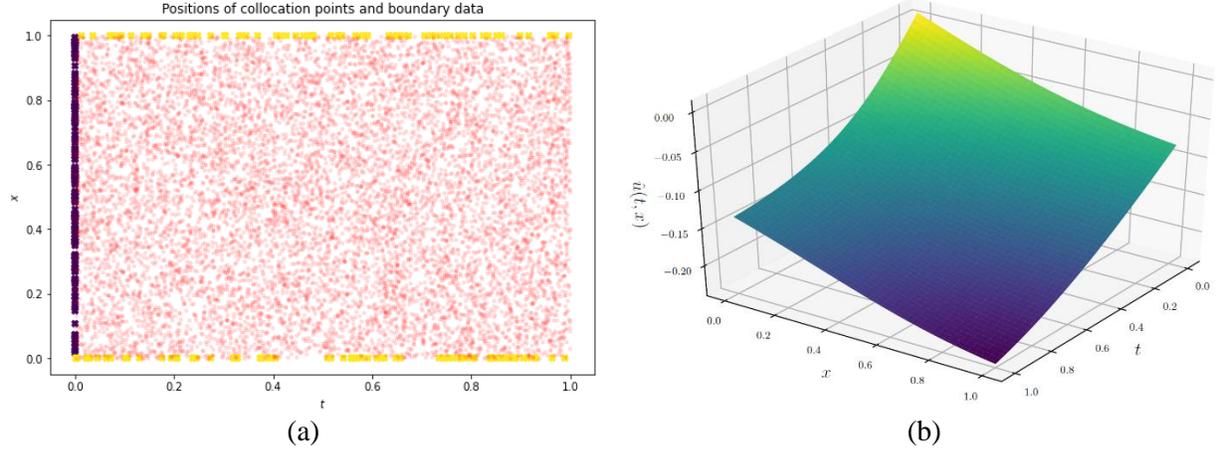

**Figure 3**: (a) Distribution of initial, boundary, and interior sampling points. (b) Initialized solution surface.

## 4. Solutions to the Buckley-Leverett Problem with a Nonconvex Flux Function using PINNs

In this section, we apply a PINN approach to solve Equation 3. In 4.1 and 4.2 we reproduce solutions from [19]. In 4.1, PINNs fail to solve the hyperbolic PDE. In 4.2 we solve Equation 3 by including a second-order derivative term in the PDE. In 4.3 we introduce the Oleinik entropy condition, and we show how to embed it in the residual loss of the PINNs approach to obtain an accurate solution to the B-L problem.

### 4.1 PINNs breakdown in the hyperbolic B-L problem

In the derivation of the B-L equation, $S_w$ is assumed to be continuous and differentiable. This assumption is necessary to arrive at the PDE in Equation 3 from the more fundamental integral laws. The solution to the problem posed by Equation 3 is triple-valued, which is not physical since it entails multiple water saturations at the same point in the domain. To correct for this unphysical solution, one might employ the equal-area rule (Welge method [35]), resulting in a discontinuity. This discontinuity is called the saturation front (shock front). The breakdown of the solution is a result of the inappropriate assumption made earlier which fails to describe the situation at the sharp front [25]. As a result of this discontinuity, the authors in [19] showed that PINNs fail to retrieve the correct solution of the hyperbolic PDE with a nonconvex flux function (Figure 4).

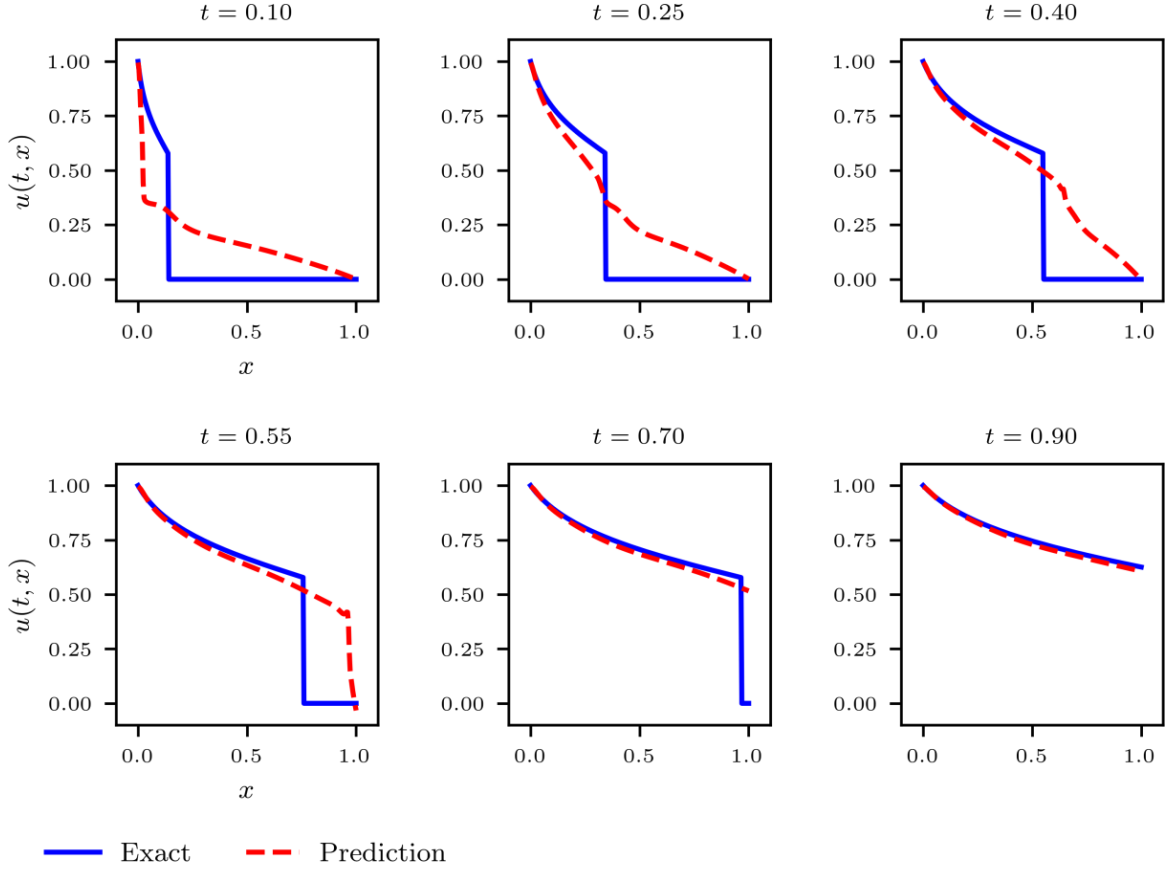

**Figure 4**: Comparison between PINNs solution (dashed line) and the analytical solution (solid line) using a nonconvex flux function and the residual in Equation 7 at five different times.

*4.2 The Vanishing-Viscosity*

A workaround for the discontinuity requires a reformulation of the PDE (Equation 3). The new PDE includes a small diffusion or a viscosity term that results in a parabolic PDE as follows:

$$\frac{\partial S_w}{\partial t} + \frac{\partial f_w(S_w)}{\partial x} = \epsilon(S_w)_{xx}, \tag{10}$$

where $\epsilon > 0$ is an extremely small number, generally in the order of $10^{-2}$ or $10^{-3}$. When $\epsilon = 0$, we retrieve the hyperbolic PDE in Equation 3. The residual becomes:

$$r := \hat{u}_t + f(\hat{u})_x - \epsilon(S_w)_{xx}. \tag{11}$$

By including a small diffusion term, one should expect a solution that closely matches the solution to the hyperbolic equation. In fact, it can be shown that the solution is unique for all time $t > 0$, for all initial conditions [27]. Solving the hyperbolic equation by including a small viscosity/diffusion term is called the vanishing-viscosity solution. In [19] it was shown that by using Equation 3 with $\epsilon = 2.5 \times 10^{-3}$, PINNs will be able to retrieve the correct solution behavior, though with slight smearing at the shock front. This behavior is to be expected since the equation is no longer hyperbolic, and the inclusion of the diffusive term smooths the shock front. The PINNs solution using the residual in Equation 11 is shown in Figure 5.

Moreover, the authors also showed that $2.5 \times 10^{-3}$ is the lower limit of $\epsilon$, and using a lower value would retrieve the hyperbolic behavior and PINNs would fail to solve the PDE. It was noted that this behavior of PINNs is synonymous with the behavior of finite-volume methods [19]. It is worth noting here that the inclusion of a second-order derivative significantly slows down PINNs (almost doubling the run time). This is because $S_w$ is a NN and computing its first derivative effectively amounts to doubling the network in size, and computing its second derivative effectively doubles the network size again.

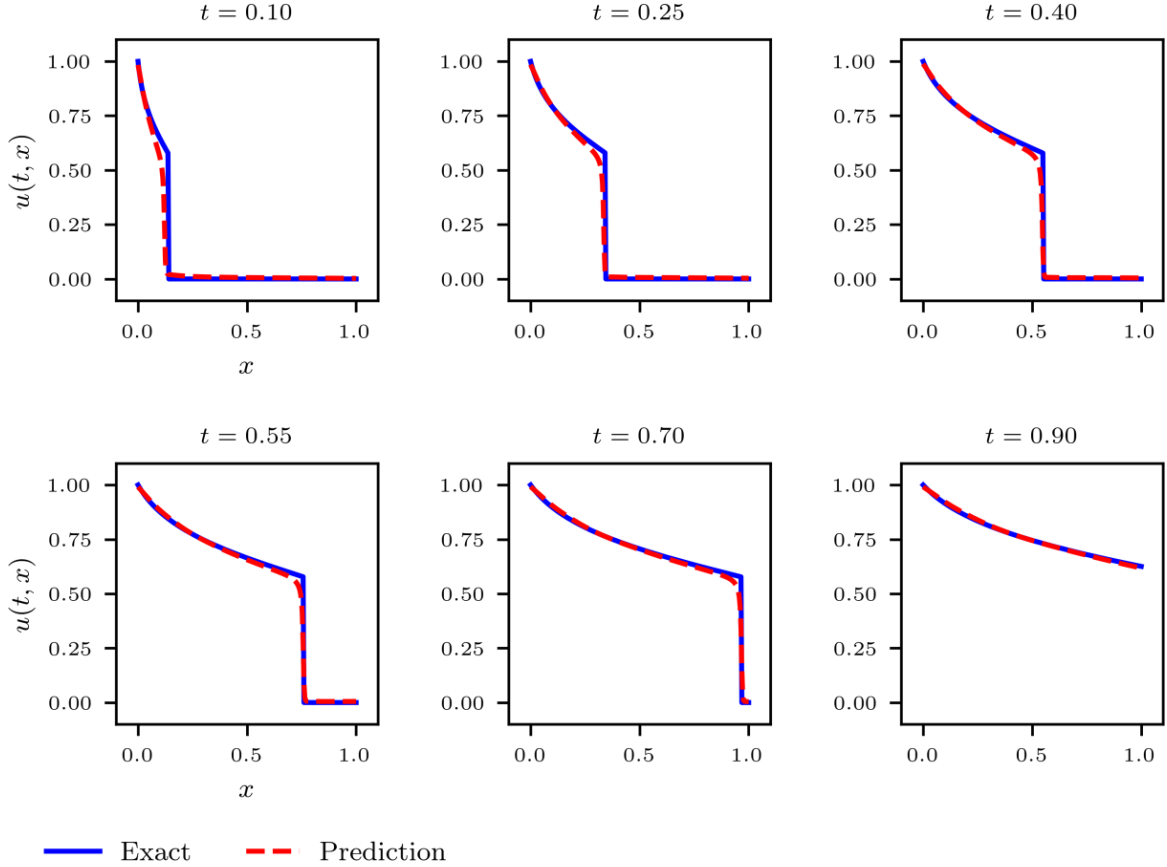

**Figure 5**: Comparison between PINNs solution (dashed line) and the analytical solution (solid line) using a nonconvex flux function and the residual in Equation 11 at five different times.

### 4.3 Solution by Constructing a Convex-Hull with RAR

Another workaround to the unphysical behavior is by using an admissibility criterion for a propagating discontinuity. For nonconvex scalar conservation laws, the Oleinik entropy condition provides such a criterion (see [26] and [27]). Oleinik entropy condition is defined as follows:

$$\frac{f(S) - f(S_l)}{S - S_l} \geq \sigma \geq \frac{f(S) - f(S_r)}{S - S_r}, \tag{12}$$

where $S_{l,r}$ are saturation values just before and after the discontinuity and $\sigma$ is the shock speed defined as:

$$\sigma = f'(S^*) = \frac{f(S^*)}{S^*}, \tag{13}$$

where $S^*$ is the saturation at the shock front. Equation 12 states that wave velocities just before the shock must be greater or equal to the shock velocity, and wave velocities just after the shock must be less than or equal to the shock velocity. A shock that obeys the entropy condition is called a self-sharpening shock. This condition maintains that $f''(S) < 0$ for all $S \in [S_r, S_l]$, meaning that $f_w(S_w)$ is now a concave function consisting of a linear portion up to $(S^*, f(S^*))$ followed by an upper concave envelope given by $f_w(S_w)$ (Equation 4). The authors in [20] were the first to suggest replacing the fractional flow equation (Equation 4) in the PINN approach with an equation that describes the convex hull. Here we suggest a slightly different equation for the convex hull given the aforementioned Oleinik entropy condition:

$$\tilde{f}_w(S_w) = \begin{cases} \sigma S_w & 0 \leq S_w < S^* \\ f_w(S_w) & S_w \geq S^* \end{cases}, \tag{14}$$

where the first part of Equation 14 describes the linear portion of the convex hull up to $(S^*, f(S^*))$, and the second part describes the concave portion given by Equation 4. We replace Equation 11 with Equation 14 in the residual of the NN. This is done by simply rewriting Equation 11 in terms of a Heaviside function. The function is then differentiated by AD as outlined earlier.

We show that the Olenik condition is sufficient to accurately solve the hyperbolic B-L problem with a nonconvex flux function. The new residual becomes:

$$r := \hat{u}_t + \tilde{f}_w(\hat{u})_x. \tag{15}$$

Here, we optimize the loss function using two different optimizers with RAR, each with its advantages and disadvantages. First, we minimize the loss function with the L-BFGS optimizer with RAR. The $\mathbb{L}_2$ validation error is $\sim 4 \times 10^{-2}$ and the loss (as per Equation 8) plateaus at $\sim 1 \times 10^{-5}$ (see orange curve in Figure 8). Second, we minimize the loss function using the Adam optimizer with RAR. This results in an improvement in the $\mathbb{L}_2$ validation error ($\sim 2 \times 10^{-2}$) and in the $\mathbb{L}_2$ loss ($\sim 1 \times 10^{-6}$) (see blue curve in Figure 8). The PINN along with the analytical solutions are shown in Figure 6, and a 3D visualization of the B-L solution using the PINN is shown in Figure 7. It can be seen from Figure 8 that the L-BFGS optimizer converges in fewer number of iterations ($\sim 2,800$) after which RAR does not result in any noticeable improvement in the loss or the validation error, suggesting that the L-BFGS optimizer gets stuck in a local minimum. Conversely, the Adam optimizer with RAR continues to reduce the loss, as evident in Figure 8, and improves the validation error; it is even possible to reduce the loss using this method an additional order of magnitude with thousands of additional iterations. The choice between the L-BFGS optimizer without RAR and the ADAM optimizer with RAR represents a tradeoff between computational efficiency and accuracy. While it is important to obtain a good solution in an efficient manner for some applications, it is equally important to obtain an accurate solution in others. The significance of an ultra-low loss, which is mainly due to the reduction of the residual loss, lies in the conservation of mass and momentum of such physics based problems in which conservation is bounded by the residual loss [3]. We note here that the Adam optimizer failed to achieve a good solution without RAR. Finally, these results clearly show that our solution method solves the pure hyperbolic form of the B-L problem.

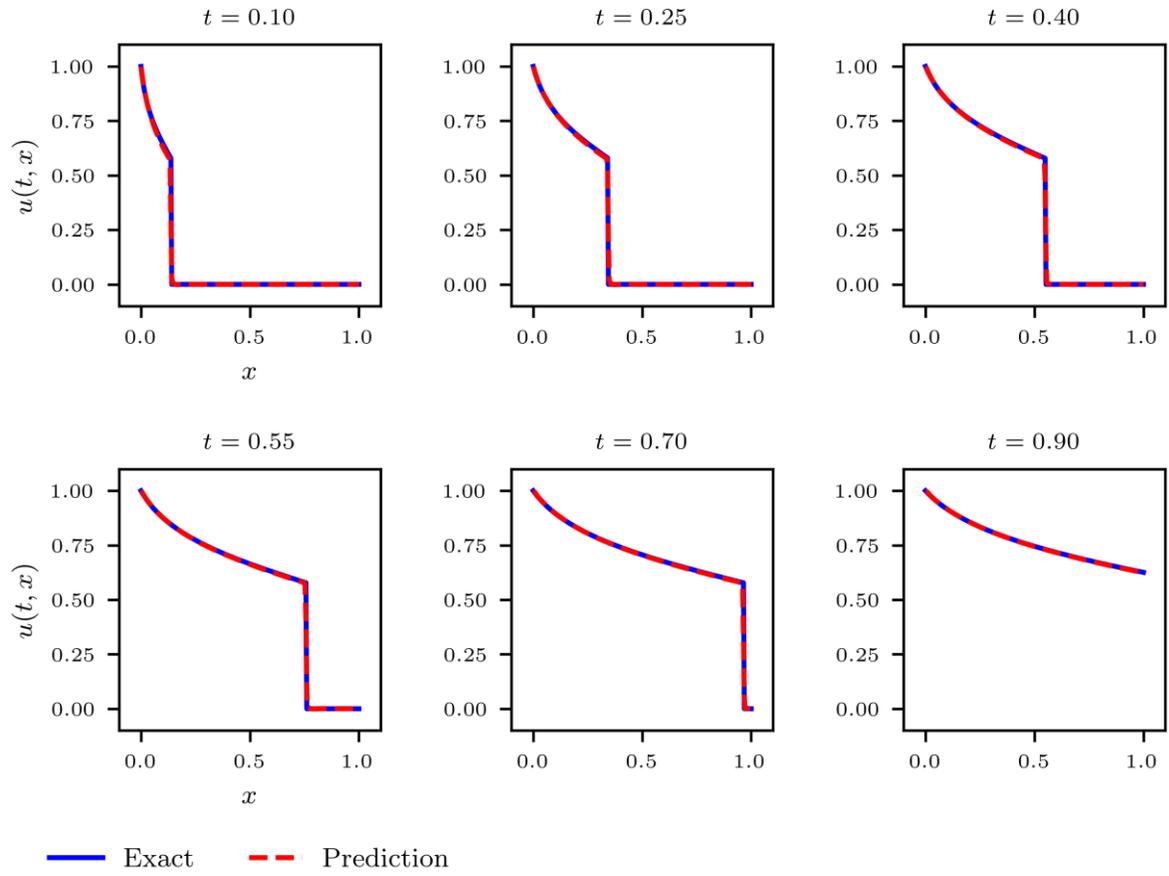

**Figure 6**: Comparison between PINNs solution (dashed line) and the analytical solution (solid line) using a nonconvex flux function and the residual in Equation 15 at five different times.

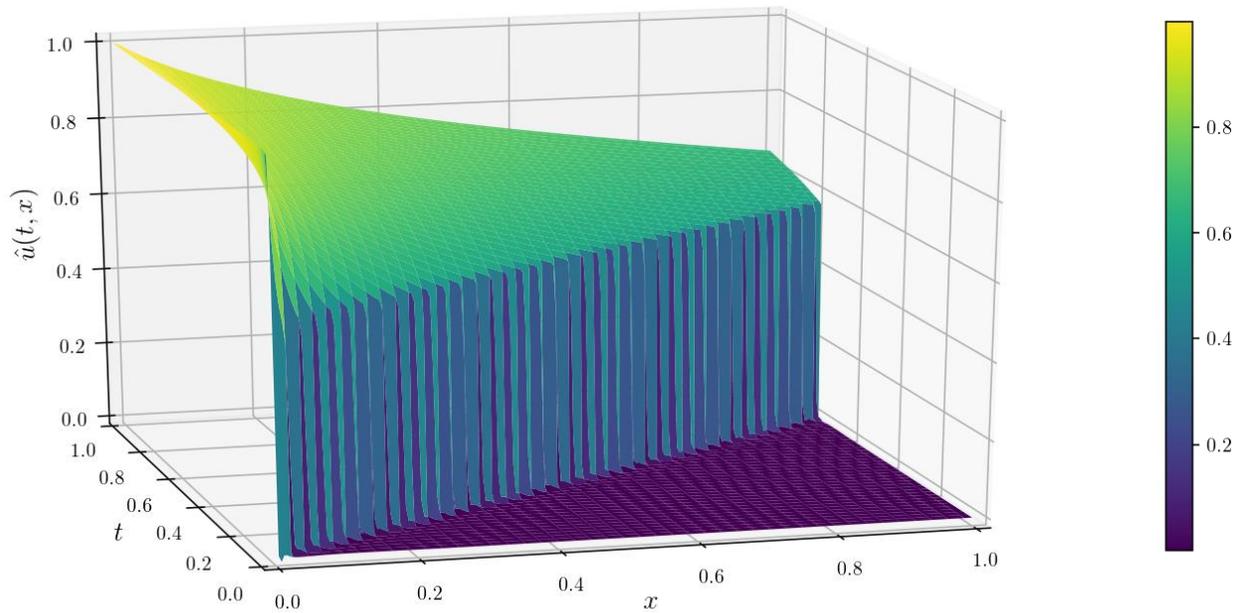

**Figure 7**: 3D visualization of the B-L solution using a nonconvex flux function and the residual in Equation 15.

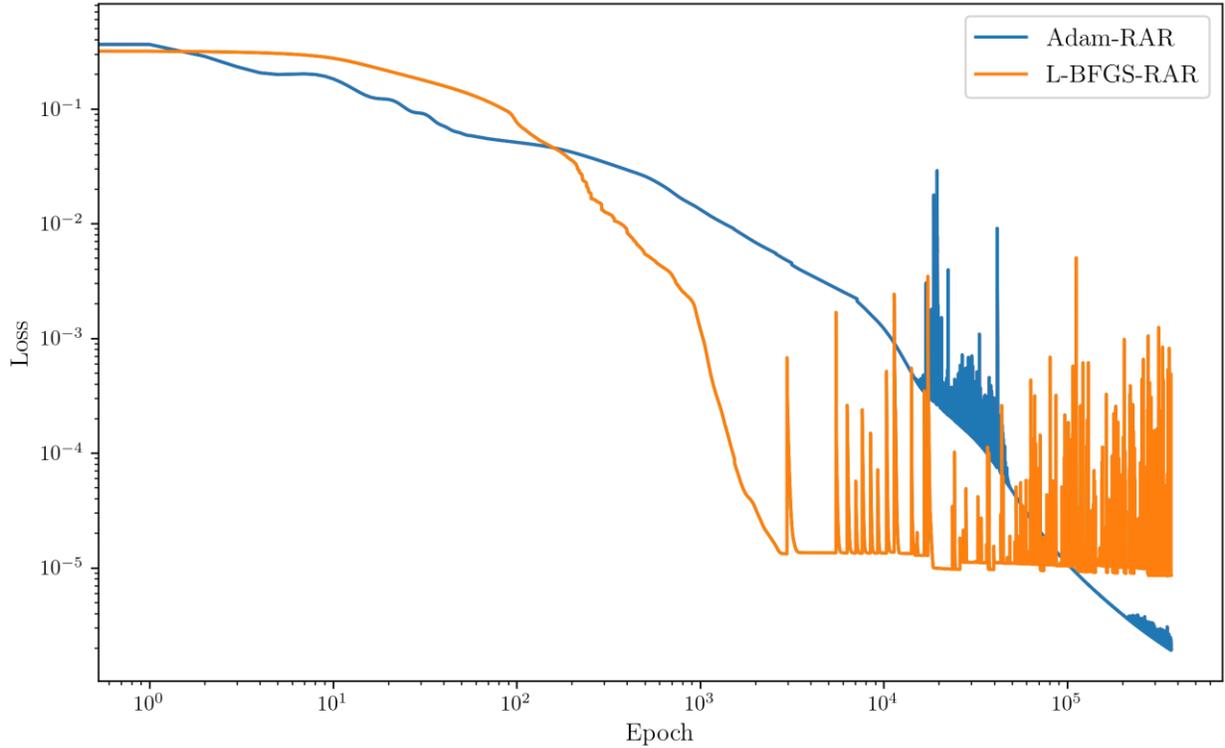

**Figure 8:** Loss of the L-BFGS optimizer (orange line), and the Adam optimizer with RAR (blue).

## 5. Conclusions

We solve the hyperbolic form of the PDE describing the B-L problem with a nonconvex flux function using PINNs. This PDE is known to be one of the most challenging to solve using traditional numerical methods (i.e. finite- difference and volume methods) in fluid flow in porous media. This goes to show the potential of PINNs in scientific computing where it can complement areas where numerical methods struggle. Our results were achieved by encoding the Oleinik entropy condition into the neural network residual. By utilizing the Adam optimization algorithm along with the RAR algorithm, an ultra-low loss and an accurate solution were achieved. Our solution is markedly better than what was previously reported in the literature on the same problem even without weighting of the loss components. We also remark here that weighting can lead to a lower loss, naturally, despite having no improvement in the validation error. These findings along with what has been presented in the literature highlight the capabilities of PINNs in solving hyperbolic PDEs given that sufficient physics is encoded in the residual. Future work in this emerging and fast-paced field will surely bring further developments and improvements to the solution of PDEs. We intend to extend the current success to more complicated problems in fluid flow in porous media. Specifically, we will focus on 2D and 3D extensions of the problem of fluid flow in porous media with added attention to heterogeneous domains. .